\documentclass[11pt,twoside]{article}

\usepackage{asp2014}

\aspSuppressVolSlug
\resetcounters

\begin{document}

\title{Cutting the cost of pulsar astronomy: Saving time and energy when searching for binary pulsars using NVIDIA GPUs}

\author{Jack~White, Karel~Ad\'amek, Wes Armour}
\affil{Oxford e-Research Centre, Department of Engineering Science, University of Oxford, 7 Keble Road, Oxford OX1 3QG, United Kingdom; \email{wes.armour@eng.ox.ac.uk}}

\paperauthor{Jack White}{jack.white@eng.ox.ac.uk}{0000-0003-2690-6858}{University of Oxford}{Oxford e-Research Centre, Department of Engineering Science}{Oxford}{Oxfordshire}{OX1 3PJ}{United Kingdom}
\paperauthor{Karel Ad\'amek}{karel.adamek@eng.ox.ac.uk}{0000-0003-2797-0595}{University of Oxford}{Oxford e-Research Centre, Department of Engineering Science}{Oxford}{Oxfordshire}{OX1 3PJ}{United Kingdom}
\paperauthor{Wes Armour}{wes.armour@eng.ox.ac.uk}{0000-0003-1756-3064}{University of Oxford}{Oxford e-Research Centre, Department of Engineering Science}{Oxford}{Oxfordshire}{OX1 3PJ}{United Kingdom}


\resetcounters

\begin{abstract}

Using the Fourier Domain Acceleration Search (FDAS) method to  search for binary pulsars is a computationally costly process. Next generation radio telescopes will have to perform FDAS in real time, as data volumes are too large to store. FDAS is a matched filtering approach for searching time-domain radio astronomy datasets for the signatures of binary pulsars with approximately linear acceleration. In this paper we will explore how we have reduced the energy cost of an SKA-like implementation of FDAS in AstroAccelerate, utilising a combination of mixed-precision computing and dynamic frequency scaling on NVIDIA GPUs. Combining the two approaches, we have managed to save 58\% of the overall energy cost of FDAS with a (<3\%) sacrifice in numerical sensitivity. 
\end{abstract}

\section{Introduction}

Software tools such as PRESTO \citep{presto} and AstroAccelerate \citep{astroaccelerate} exist to search for the signatures of binary pulsar systems in time-domain radio astronomy datasets. As these tools implement FDAS, efforts have been made during their development to minimise the execution time of this approach \citep{Dimoudi_2018} \citep{adamek2020gpu}. Since datacentre-scale computing facilities will be required \citep{levin2017pulsar} to perform FDAS at the scale required for next generation radio telescopes, the focus of this work is on reducing the energy cost of FDAS, rather than solely execution time. The ultimate goal is therefore to reduce the CO$_2$ emissions associated with generating the power required to run these facilities.

\section{Methodology}

\subsection{FDAS}
As previously mentioned, the Fourier Domain Acceleration Search is a method for searching for binary pulsars with approximately linear acceleration \citep{Ransom2002:FDAS}. The linearity requirement constrains the observation time ($T_{obs}$) to be less than  $\approx10\%$ of the orbital period ($P_{orb}$) of the pulsar and its orbital companion. As it is a matched filtering based approach, the computational cost of running FDAS scales linearly with the number of acceleration values the astronomer searches for.

\subsection{Mixed-Precision}

FDAS relies on the Fast Fourier Transform (FFT) to perform convolutions in the frequency domain. When implemented on GPUs, the performance of an FFT is limited by available memory bandwidth \citep{adamekfft}. This motivates the desire to reduce the memory footprint of the dataset being processed. One way to achieve this is by reducing the floating point precision of each number. Initially, AstroAccelerate used IEEE-754 single-precision 32-bit floating point numbers for the entire pipeline \citep{dimoudi2015pulsar}, \citep{adamek2017improved}. In previous work \citep{bfloatarxiv}, we converted the most computationally demanding section (the convolution stage) of the pipeline to use bfloat16. This is a 16-bit floating point format proposed by Google Brain, which is supported on Ampere (and newer) architecture NVIDIA GPUs. The numerical error introduced by converting the convolution section of the FDAS pipeline to bfloat16 resulted in a maximum error of 2.7\% in the magnitude of detection strength.

\subsection{Dynamic Frequency Scaling}

The performance of FDAS is limited by available memory bandwidth rather than the computational throughput of the GPU. By default, during program execution NVIDIA GPUs run at the highest available core clock frequency (within thermal constraints). Therefore, if the GPU core is not fed with sufficient data to saturate its computational throughput, it could be wasting energy while waiting for data to arrive. We reduced the clock frequency of the GPU core during the execution of FDAS, and subsequently were able to observe a reduction in overall energy usage, despite an increase in execution time.

\section{Results}

\begin{figure}[h!]
\includegraphics[width=1\textwidth]{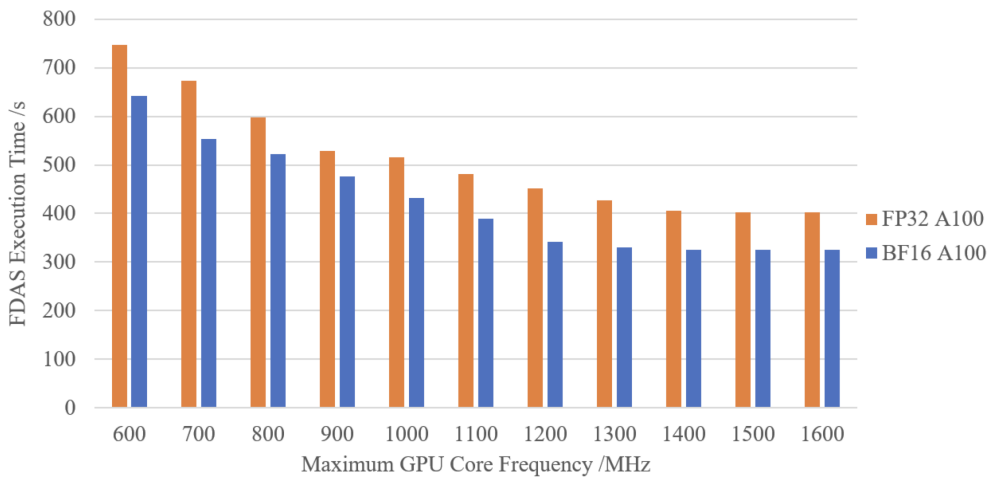}
\caption{The execution time improvement associated with introducing mixed-precision and dynamic frequency scaling. Data collected using an NVIDIA A100 GPU.}
\label{fig:executiontime}
\end{figure}

Figure \ref{fig:executiontime} shows the execution time of FDAS when run in an SKA-like configuration ($600$ second $T_{obs}$, $64 \mu s$ resolution, 5925 DM trials with 3050 at full time resolution, 193 acceleration trials) using an NVIDIA A100 GPU (SXM4). The notable aspect of this graph is how there are diminishing returns in execution time improvement at the highest frequencies, demonstrating that the pipeline is bandwidth bound.

\begin{figure}[h!]
\includegraphics[width=1\textwidth]{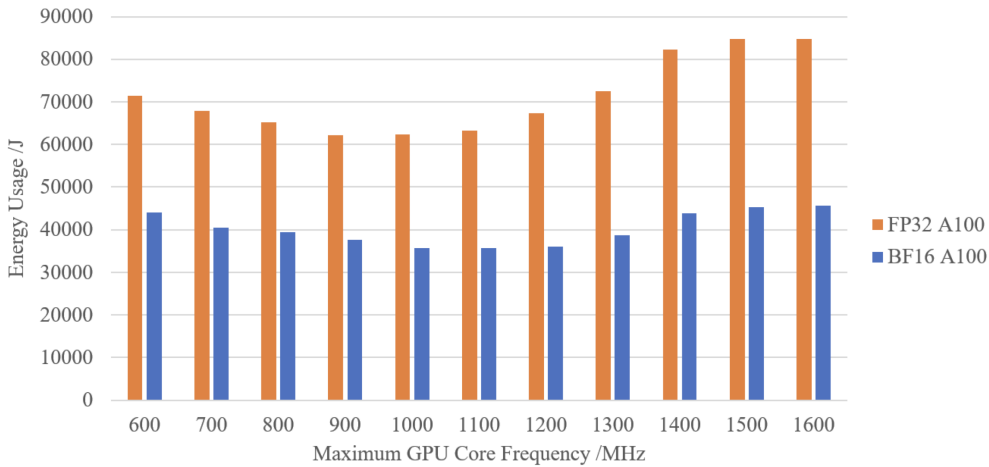}
\caption{The energy saving associated with introducing mixed-precision and dynamic frequency scaling. Data collected using an NVIDIA A100 GPU.}
\label{fig:energyusage}
\end{figure}

Figure \ref{fig:energyusage} shows the energy usage of the GPU while performing FDAS in an SKA-like configuration at a range of GPU core frequencies. Notably, the lowest energy usage does not coincide with the shortest execution time. This deviates from the assumption that the fastest version of code is usually the most efficient, and demonstrates that when the core is underclocked, the GPU is drawing significantly less power.

\begin{figure}[h!]
\includegraphics[width=1\textwidth]{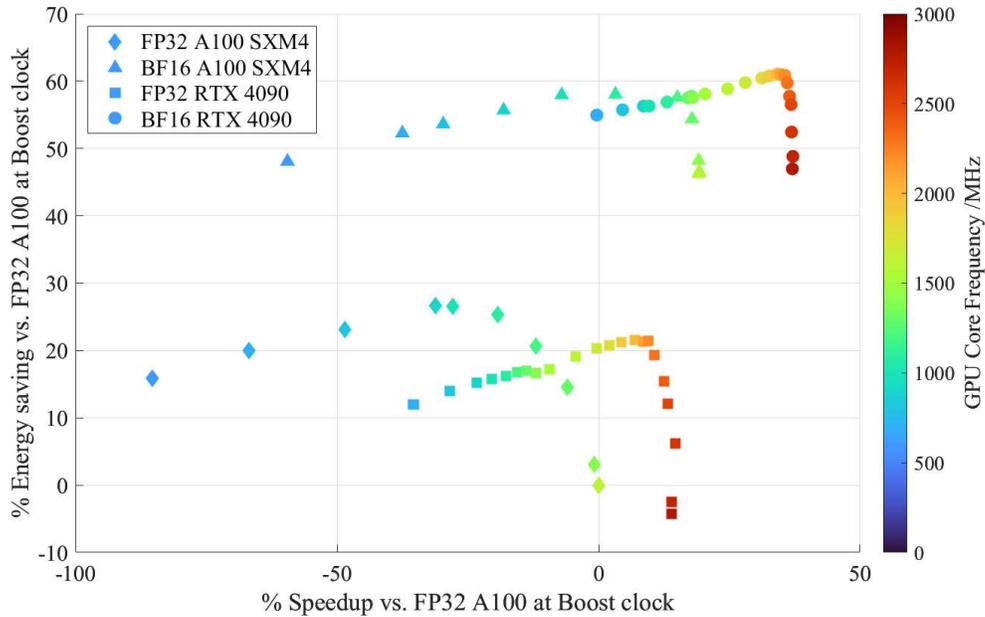}
\caption{Percentage energy saving compared against percentage speedup associated with introducing mixed-precision and dynamic frequency scaling on different architecture GPUs. An overall optimal configuration maximises both axes.}
\label{fig:scatterplot}
\end{figure}

By combining the two approaches (mixed-precision and dynamic frequency scaling), we were able to achieve a 58\% reduction in the energy usage of performing FDAS in an SKA-like configuration. This data was collected using an A100 GPU, and is compared to the single-precision implementation without dynamic frequency scaling. 

Figure \ref{fig:scatterplot} combines the axes of Figures \ref{fig:executiontime} and \ref{fig:energyusage} to highlight the overall optimal configuration, and includes data collected using an RTX 4090. This primarily highlights the speedup associated with using a newer generation of hardware, although the energy saving improvement is only marginal.

\section{Summary}

The goal of this work was to investigate the impact of combining dynamic frequency scaling and mixed-precision on the energy usage of FDAS. The results we have presented show that reducing core clock frequency consistently increases execution time. However, it also disproportionately decreases the power drawn by the GPU, which allows large energy savings to be made.

Combining dynamic frequency scaling with mixed-precision allows the resulting implementation to be significantly more efficient, while remaining faster than the unmodified implementation.

\bibliography{C16}


\end{document}